\documentclass[onecolumn]{aastex63}
\usepackage{amsmath}
\usepackage{csvsimple}
\usepackage{graphicx}
\usepackage{color}
\usepackage{xcolor}

\shorttitle{\cnce\ RM Detection}
\shortauthors{Zhao, Kunovac, et al.}

\newcommand{\acronym}[1]{{\small{#1}}}
\newcommand{\code}[1]{\texttt{#1}}
\newcommand{\expres}{\acronym{EXPRES}}
\newcommand{\tess}{\acronym{TESS}}
\newcommand{\kepler}{\acronym{Kepler}}
\newcommand{\ldt}{\acronym{LDT}}

\newcommand{\espresso}{\acronym{ESPRESSO}}

\newcommand{\harps}{\acronym{HARPS}}

\newcommand{\sme}{\acronym{SME}}
\newcommand{\lightkurve}{\code{lightkurve}}
\newcommand{\exoplanet}{\code{exoplanet}}
\newcommand{\starry}{\code{starry}}
\newcommand{\pymc}{\code{PyMC3}}
\newcommand{\theano}{\code{Theano}}
\newcommand{\ldtk}{\code{LDTk}}

\newcommand{\ellc}{\code{ellc}}
\newcommand{\emcee}{\code{emcee}}
\newcommand{\celerite}{\code{celerite}}
\newcommand{\phoenix}{\code{PHOENIX}}
\newcommand{\cnc}{55~Cnc}
\newcommand{\cnce}{55~Cnc~e}
\newcommand{\cms}{\mbox{cm s\textsuperscript{-1}}}
\newcommand{\ms}{\mbox{m s\textsuperscript{-1}}}
\newcommand{\kms}{\mbox{km s\textsuperscript{-1}}}

\newcommand{\vsini}{$v \sin i_\star$}
\newcommand{\teff}{$T_\mathrm{eff}$}

\newcommand{\logg}{log $g$}

\newcommand{\msun}{$M_{\odot}$}
\newcommand{\rsun}{$R_{\odot}$}

\newcommand{\mearth}{$M_{\oplus}$}
\newcommand{\rearth}{$R_{\oplus}$}
\newcommand{\mjup}{$M_\mathrm{Jup}$}

\begin{document}

\title{Measured Spin-Orbit Alignment of Ultra-Short Period Super-Earth 55~Cancri~e}
\correspondingauthor{Lily L. Zhao}
\email{lzhao@flatironinstitute.org}

\author[0000-0002-3852-3590]{Lily L. Zhao}
\affiliation{Center for Computational Astrophysics, Flatiron Institute, Simons Foundation, 162 Fifth Avenue, New York, NY 10010, USA} 

\author[0000-0001-9419-3736]{Vedad Kunovac} 
\affiliation{Lowell Observatory, 1400 W. Mars Hill Rd., Flagstaff, AZ 86001, USA}
\affiliation{Department of Astronomy and Planetary Science, Northern Arizona University, PO Box 6010, Flagstaff, AZ 86001, USA}

\author[0000-0002-9873-1471]{John M. Brewer}
\affiliation{Dept. of Physics \& Astronomy, San Francisco State University, 1600 Holloway Ave., San Francisco, CA 94132, USA}

\author[0000-0003-4450-0368]{Joe Llama} 
\affiliation{Lowell Observatory, 1400 W. Mars Hill Rd., Flagstaff, AZ 86001, USA}

\author[0000-0003-3130-2282]{Sarah C. Millholland} 
\affiliation{MIT Kavli Institute for Astrophysics and Space Research, Massachusetts Institute of Technology, Cambridge, MA 02139, USA}

\author[0000-0002-3385-8391]{Christina Hedges}
\affiliation{University of Maryland, Baltimore County, 1000 Hilltop Cir, Baltimore, MD 21250, USA} 
\affiliation{NASA Goddard Space Flight Center, 8800 Greenbelt Rd, Greenbelt, MD 20771, USA}

\author[0000-0002-4974-687X]{Andrew E. Szymkowiak}
\affiliation{Department of Astronomy, Yale University, 52 Hillhouse Ave., New Haven, CT 06511, USA}
\affiliation{Department of Physics, Yale University, 217 Prospect St, New Haven, CT 06511, USA}

\author[0000-0002-9288-3482]{Rachael M. Roettenbacher} 
\affiliation{Department of Astronomy, University of Michigan,1085 S.\ University Ave., Ann Arbor, MI 48109, USA}  

\author[0000-0001-9749-6150]{Samuel H. C. Cabot}
\affiliation{Department of Astronomy, Yale University, 52 Hillhouse Ave., New Haven, CT 06511, USA}

\author[0000-0002-5870-8488]{Sam A. Weiss}
\affiliation{Department of Astronomy, Yale University, 52 Hillhouse Ave., New Haven, CT 06511, USA}

\author[0000-0003-2221-0861]{Debra A. Fischer}
\affiliation{Department of Astronomy, Yale University, 52 Hillhouse Ave., New Haven, CT 06511, USA}

\begin{abstract}
A planet's orbital alignment places important constraints on how a planet formed and consequently evolved.  The dominant formation pathway of ultra-short period planets ($P<1$ day) is particularly mysterious as such planets most likely formed further out, and it is not well understood what drove their migration inwards to their current positions.  Measuring the orbital alignment is difficult for smaller super-Earth/sub-Neptune planets, which give rise to smaller amplitude signals.  Here we present radial velocities across two transits of 55~Cancri~e, an ultra-short period Super-Earth, observed with the Extreme Precision Spectrograph (\expres).  Using the classical Rossiter-McLaughlin (RM) method, we measure \cnce's sky-projected stellar spin-orbit alignment (i.e., the projected angle between the planet's orbital axis and its host star's spin axis) to be $\lambda=10\substack{+17\\ -20}^{\circ}$ with an unprojected angle of $\psi=23\substack{+14\\ -12}^{\circ}$.  The best-fit RM model to the \expres\ data has a radial velocity semi-amplitude of just $0.41\substack{+0.09\\ -0.10}$ \ms.  The spin-orbit alignment of \cnce\ favors dynamically gentle migration theories for ultra-short period planets, namely tidal dissipation through low-eccentricity planet-planet interactions and/or planetary obliquity tides.
\end{abstract}

\section{Introduction}

55~Cancri hosts five known exoplanets with minimum mass estimates ranging from approximately 8 \mearth\ to 3 \mjup\ and periods less than one day to nearly 20 years \citep{butler1997, marcy2002, mcarthur2004, fischer2008}.  Of particular interest has been \cnce, one of the most massive known ultra-short-period planets (USP) and the only planet around \cnc\ found to transit \citep{winn2011, demory2011}.  It has an orbital period of $0.7365474\substack{+1.3E-6\\ -1.4E-6}$ days, a mass of 7.99$\pm$0.33 \mearth, and a radius of $1.853\substack{+0.026\\ -0.027}$ \citep{dawson2010, bourrier2018}.  A precise measure of \cnce's stellar spin-orbit alignment---the angle between the host star's spin axis and the planet's orbit normal---will shed light on the formation and evolution of USPs, especially in the case of compact, multi-planet systems.

It has been shown that USPs form a statistically distinct population of planets \citep{steffen2016} that tend to be misaligned with other planetary orbits in their system \citep{dai2018}.  This suggests that USPs experience a unique migration pathway that brings them close in to their host stars. This inward migration is most likely driven by dissipation due to star-planet tidal interactions that result from either non-zero eccentricities \citep{petrovich2019, pu2019} or planetary spin-axis tilts \citep{millholland2020}.

The stellar spin-orbit alignment of a transiting planet can be measured by obtaining spectroscopic observations during the planet's transit.  Commonly called the Rossiter-McLaughlin (RM) effect \citep{triaud2018, albrecht2022}, radial velocity (RV) deviations as a planet transits a rotating star reveal whether the planet is transiting across the blue-shifted half of the star that rotates towards the observer or the red-shifted half of the star that rotates away.  Capturing the resultant net red-/blue-shift reveals the orientation of the planet's orbital normal vector with respect to its host star's spin vector, i.e., the sky-projected stellar spin-orbit alignment or the stellar obliquity.

\begin{table}[b!]
\scriptsize
\caption{\cnce\ Fit Parameters}
\label{tab:transitFit}
\begin{center}
\begin{tabular}{l c c l l}
\hline
\hline
Parameter & Symbol & Value & Unit & Prior \\
\hline
\textbf{Transit Fit} & & & & \\
Planet/star radii ratio & $R_\mathrm{p}/R_{\star}$ & $0.018016\substack{+0.000085\\ -0.000077}$ & & LogNormal$(-8, 10)$ \\ 
Radius & $R_p$ & $1.856\substack{+0.024\\ -0.025}$ & \rearth & -- \\ 
Orbital period & $P$ & $0.7365430 \pm 0.0000014$ & days &  Normal$(0.736546, 1)$ \\ 
Transit epoch & $T_0$ & $2459511.487987\substack{+0.000071\\ -0.000075}$ & BJD$_\mathrm{TDB}$ & Normal$(2459511.48778, 1)$ \\ 
Impact parameter & $b$ & $0.373\substack{+0.027\\ -0.032}$ & $R_\star$ &  Uniform$(0, 1)$ \\ 
Orbital inclination & $i$ & $83.9\substack{+0.6\\ -0.5}$ & deg & -- \\
Orbital separation & $a/R_\star$ & $3.516\substack{+0.042\\ -0.041}$ & & -- \\
Transit duration & $T_{14}$ & $1.5439\substack{+0.0027\\ -0.0026}$ & hours & -- \\ 
Limb-darkening & $u_1$ & 0.4823$\pm$0.0017 & & Normal$(0.4824, 0.0017)$ \\ 
 & $u_2$ &$0.1276\substack{+0.0037\\ -0.0041}$ & & Normal$(0.1280, 0.0038)$ \\ 
\hline
\textbf{RM Fit} & & & & \\
\textit{Fitted parameters} \\
Sky-projected obliquity & $\lambda$ & $10\substack{+17\\ -20}$  & deg & Uniform$(-180^\circ, 180^\circ)$ \\ 
Projected rotational velocity & $v\sin{i_\star}$ & $2.00\substack{+0.43\\ -0.47}$ & \kms & Uniform$(0, 18)$ \\
Net convective blueshift velocity & $V_\mathrm{CB}$ & $-157\substack{+86\\-94}$ & \ms & Uniform$(-1000, 0)$, Normal$(-150, 100)$ \\ [2pt]
\textit{Derived parameters} \\
RM semi-amplitude & $K_\mathrm{RM}$ & $0.41\substack{+0.09\\ -0.10}$ & \ms \\
Stellar inclination & $i_\star$ & $75\substack{+11\\ -18}$ & deg \\
Obliquity & $\psi$ & $23\substack{+15\\ -12}$ & deg \\
\hline
\end{tabular}
\end{center}
\end{table}

\begin{figure}[thb]
\centering
\includegraphics[width=0.85\textwidth]{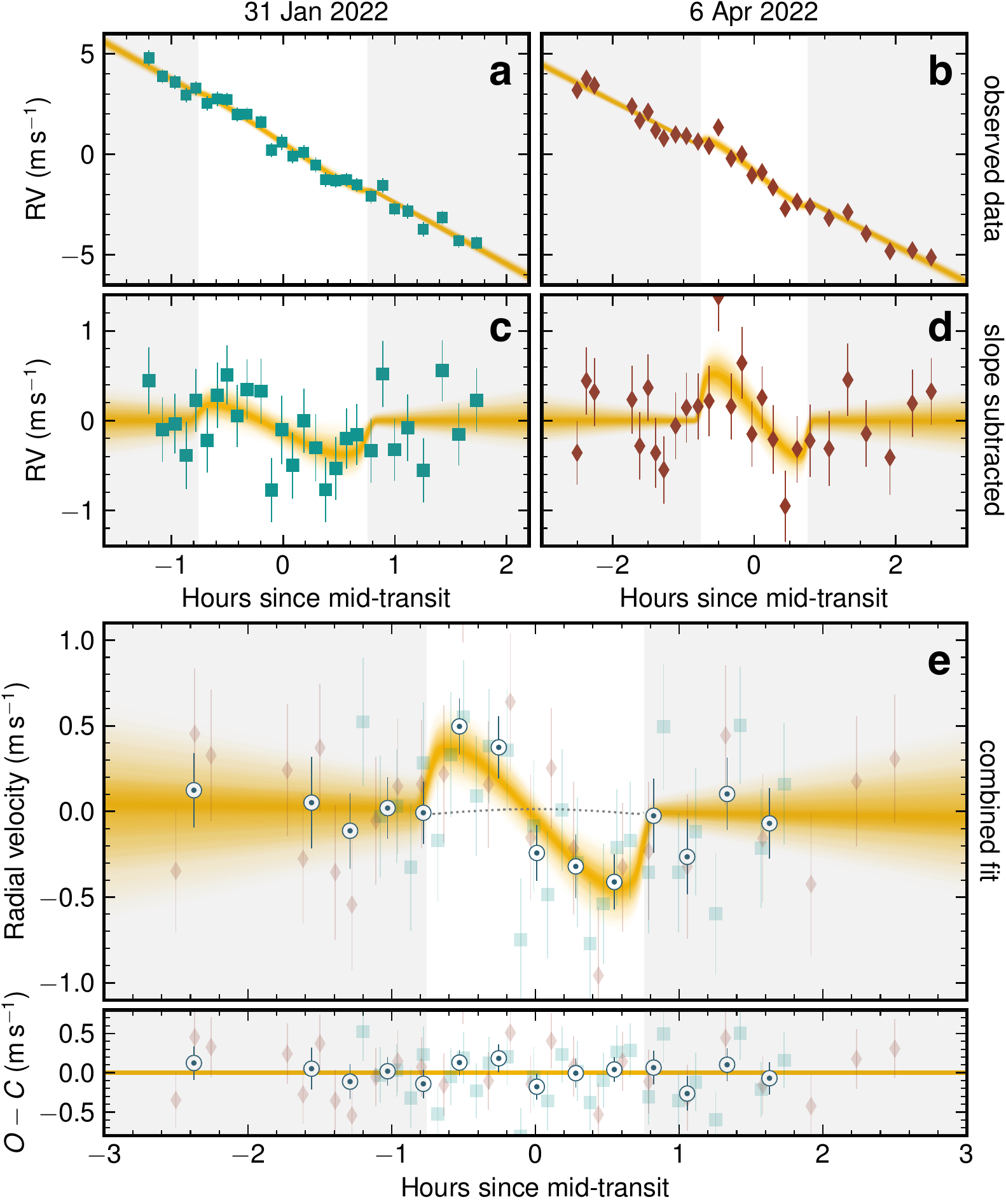}
\caption{Rossiter-Mclaughlin model fit to two nights of \expres\ RV data.  Data from Jan.\ 31, 2022 are shown as green squares, Apr.\ 6, 2022 as red diamonds.  The RV value (indicated by each marker) represents the best-fit Doppler shift produced by the \expres\ chunk-by-chunk, forward modeling pipeline with error bars spanning the corresponding 1-$\sigma$ errors (see \citep{petersburg2020} for more information on the \expres\ pipeline).  RM models are shown as yellow curves.  Shaded yellow regions span 50\textsuperscript{th} to 99\textsuperscript{th} percentile of the posterior sampled by the MCMC. The white highlighted area spans the transit duration of \cnce.  Panels a-d show each of the two nights of observations individually.  Panels a and b show the original RVs along with the full model fit for each night.  Panels c and d show the RVs and model fit with the best fit linear slope removed.  Panel e shows the combined fit to both nights of data and the residuals to the best fit.  White circular points represent both nights of data binned by phase with points every 0.015 phase units (${\sim}~15.9$ minutes) to help guide the eye.  Phase binned points are calculated as the average of all data points within the phase bin weighted by their errors, $\sigma_i$, with associated errors given by $\sqrt{\frac{1}{\sum 1/\sigma^2_i}}$.  The dotted gray line represents the center-to-limb convective blueshift contribution to the RVs, which has a semi-amplitude of 1.3 \cms.}
\label{fig:rm_fit}
\end{figure}

Previous attempts to measure the RM effect for \cnce\ have produced mixed results.  Using data from the High Accuracy Radial velocity Planet Searcher for the Northern hemisphere (\harps-N), \cite{bourrier2014} reported a sky-projected obliquity of $\lambda = 72\substack{+13 \\ -12}^{\circ}$.  However, the best fit model from that analysis also returned a stellar projected rotational velocity of \vsini = 3.3$\pm$0.9 \kms, which is not consistent with the \vsini\ measured by other studies \citep{bourrier2018}.  A separate observing campaign also using \harps-N data reported a non-detection and rule out an RM signal with an amplitude greater than $\sim$35 \cms\ \citep{lopezmorales2014}.

Here we present a $41\substack{+9\\ -10}$ \cms\ amplitude RM detection of \cnce\ using observations from the Extreme Precision Spectrograph (\expres).  The arrival of ultra-stabilized spectrographs, including \expres\ and \espresso\ among others, have made sub-meter-per-second RV precision possible \citep{pepe2013, jurgenson2016}.  This measurement is the smallest amplitude classic RM detection to date.  Precision RV measurements from spectrographs like \expres\ will be able to detect lower-amplitude RM effects that have previously been missed in the regime of small planets or slowly rotating stars.

\section{Results}

We use two sectors of \tess\ photometry to fit for the transit parameters of \cnce, which are then used to constrain the RM model.  We use RV measurements from \expres\ to detect and fit for the RM effect during two \cnce\ transits observed on Jan.\ 31 2022 and Apr.\ 6, 2022.  The \expres\ spectra were also used to derive stellar properties for \cnc\ using Spectroscopy Made Easy (SME) \citep{brewer2016, piskunov2017}.

The \expres\ RVs and RM fits are shown in Figure~1.  The best-fit RM models for each individual night as well as the combined data are plotted over the \expres\ data in yellow with darker and lighter shading showing the 50\textsuperscript{th} and 90\textsuperscript{th} percentile of the posterior distribution, respectively.  The full posterior distribution was sampled using a Markov chain Monte Carlo (MCMC). 

With the combined fit using both nights of observations, we find that \cnce\ has a sky-projected stellar obliquity of $\lambda=10\substack{+17 \\ -20}^{\circ}$.  By incorporating the rotation period of \cnc\ \citep{bourrier2018} we derive an un-projected stellar obliquity of $\psi=23\substack{+14 \\ -12}^{\circ}$.  We therefore find that \cnce's orbit normal is close to aligned with its host star's rotation axis.  The best-fit transit and RM parameters are given in Table~\ref{tab:transitFit} along with the priors used in the fitting.  Other nuisance parameters needed for each fit as well as the results of the fits to each individual night are given in Table~\ref{tab:fit_nuisance_params} and described more fully in Section~\ref{sec:methods} below.  The measured stellar spin-orbit alignment of \cnce\ is graphically represented in Figure~2.

\begin{figure}[htp]
\centering
\includegraphics[width=0.5\textwidth]{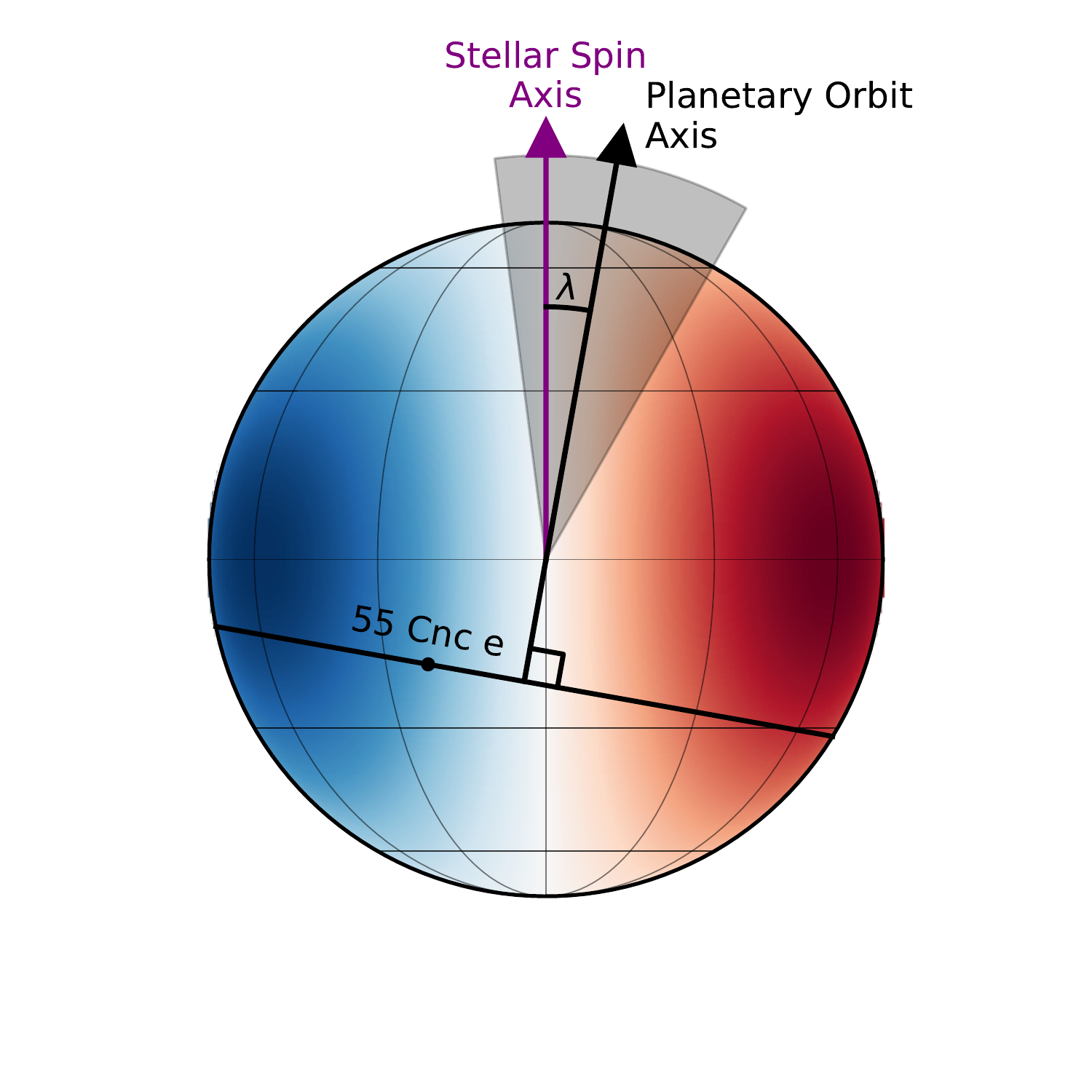}
\caption{A visual representation of the sky-projected alignment between \cnce's stellar spin axis (purple) and planetary orbit axis (mean shown in black).  The apparent Doppler shift of the star's surface that arises from its rotation is shown in shades of blue to red, which correspond to the degree to which the stellar surface appears blue- or red-shifted.  Lighter shading corresponds to smaller Doppler shift.  The star is shown with an inclination $i_{\star}=90^{\circ}$, which is consistent with the wide constraints on our estimate of $i_{\star}$. The 1-$\sigma$ confidence level for \cnce's orbit axis is shown by the gray shading.  The transit trajectory is shown as a black band. The size of the planet is to scale with the size of the star.}
\label{fig:alignment}
\end{figure}

\begin{table}[bth]
\scriptsize
\caption{\tess\ and \expres\ Fitted Nuisance Parameters}
\label{tab:fit_nuisance_params}
\begin{center}
\begin{tabular}{l c c l l}
\hline
\hline
Parameter & Symbol & Value & Unit & Prior\\
\hline
\textbf{\tess\ Fit} & & & \\
\textit{Sector 44} & & & \\
GP amplitude & $\sigma_\mathrm{GP}$ & $0.234\substack{+0.028\\ -0.022}$ & ppt & LogNormal$(-8, 10)$  \\ 
GP timescale & $\tau_\mathrm{GP}$ & $1.086\substack{+0.135\\ -0.105}$ & day &  LogNormal$(2, 10)$ \\
Mean flux & $f$ &  $0.045 \pm 0.035$ & ppt & Normal$(0, 10)$ \\
White noise & $\sigma$ &  $0.0980 \pm 0.0012$  & ppt & LogNormal$(-16, 10)$ \\[2pt]
\textit{Sector 46} & & & \\
GP amplitude & $\sigma_\mathrm{GP}$ & $0.112\substack{+0.016\\ -0.013}$ & ppt & LogNormal$(-8, 10)$ \\ 
GP timescale & $\tau_\mathrm{GP}$ & $2.22\substack{+0.32\\ -0.23}$ & day & LogNormal$(2, 10)$ \\ 
Mean flux & $f$ &  $0.018\substack{+0.022\\ -0.021}$ & ppt & Normal$(0, 10)$ \\
White noise & $\sigma$ & $0.0965\substack{+0.0011\\ -0.0012}$ & ppt & LogNormal$(-16, 10)$ \\ [2pt]
\hline 
\textbf{\expres\ Fit} & & & \\
\textit{Jan.\ 31 2022} & & & \\
Sky-projected obliquity & $\lambda$ & $34\substack{+24\\ -26}$  & deg & $\mathrm{Uniform}(-180^\circ, 180^\circ)$ \\ 
Projected rotational velocity & $v\sin{i_\star}$ & $1.74\substack{+0.62\\ -0.68}$ & \kms & $\mathrm{Uniform}(0, 18)$ \\
Intercept & $y_0$ & $0.646\substack{+0.034\\ -0.033}$  & \ms  & -- \\ 
Slope & $m$ & $-72.4\substack{+2.0\\ -2.1}$  & $\mathrm{m}\,\mathrm{s}^{-1}\,\mathrm{day}^{-1}$ & $\mathrm{Uniform}(-150, -50)$  \\[2pt]
\textit{Apr.\ 6 2022} & & & \\
Sky-projected obliquity & $\lambda$ & $-19\substack{+30\\ -20}$  & deg & Uniform$(-180^\circ, 180^\circ)$ \\ 
Projected rotational velocity & $v\sin{i_\star}$ & $2.36\substack{+0.59\\ -0.65}$ & \kms & $\mathrm{Uniform}(0, 18)$ \\
Intercept & $y_0$  & $-0.952\substack{+0.023\\ -0.022}$  & \ms & -- \\ 
Slope & $m$ & $-43.1\substack{+1.4\\ -1.3}$  & $\mathrm{m}\,\mathrm{s}^{-1}\,\mathrm{day}^{-1}$ & $\mathrm{Uniform}(-150, -5)$ \\ 
\hline
\multicolumn{5}{l}{\footnotesize Note: The given uncertainties represent each measurement's 68\% credible interval.}
\end{tabular}
\end{center}
\end{table}

\section{Discussion} \label{sec:discussion}
We recover a statistically significant detection of the RM effect for \cnce\ using \expres\ data with a false alarm probability of 0.3\%.  The amplitude of the measured signal, $K_\mathrm{RM}=0.41\substack{+0.09\\ -0.10}$~\ms, is the smallest classically recovered RM measurement to date.  Two other super-Earth systems have similar expected RM amplitudes, but could only be detected using stellar line profile analysis \citep{kunovachodzic2021, bourrier2021}.

This result clarifies previous measurements of \cnce's stellar spin-orbit alignment, which gave contradictory results \citep{bourrier2014,lopezmorales2014}.  \cite{bourrier2014} previously reported best fit RM values of $\lambda = 72\substack{+13 \\ -12}^{\circ}$ and $v \sin i_\star = 3.3 \pm 0.9$ \kms.  This \vsini\ value is $\sim 1.8 \sigma$ greater than the literature value derived from stellar line broadening, which report the \vsini\ as $1.7 \pm 0.5$ \kms \citep{brewer2016}.  The RM fit presented here returns a best-fit \vsini\ of $2.00\substack{+0.43\\ -0.47}$ \kms, which is within $1\sigma$-agreement of the literature value.  Our RM fit additionally agrees with the independent SME line broadening analysis, which returned a \vsini\ of $1.94\pm0.50$\kms.  These \vsini\ values are in some tension with the rotational velocity calculated using the stellar radius and rotation period ($v=\frac{2\pi R_{\star}}{P_{\star}}=1.23\pm0.01$ \kms) by $\sim 1.5 \sigma$.  However, the agreement between the two methods used here as well as agreement with literature values \citep{valenti2005,brewer2016} suggests this is due to uncertainties in the quoted rotation period possibly due to differential rotation or spots.

The \expres\ data used in this analysis have a consistent and often higher SNR as well as lower uncertainties than the RV measurements previously used.  All \expres\ observations were integrated until they reached a constant SNR of 225 at 550 nm; this consistency in SNR allowed for more stable final RVs with similar error properties.  Though it is difficult to compare SNR between different spectrographs given each instrument's unique throughput response as a function of wavelength, we report that \cite{lopezmorales2014} used observations that ranged in SNR between 100 and 275 with a median SNR of 165 at 624 nm.  \cite{bourrier2014} give a median SNR across all nights of 144 at 527 nm with the median SNR within a night ranging from 92 to 354.  The \harps-N RVs previously used have a median error of about 80 \cms\ and were calibrated using either a ThAr lamp or a Fabry–P\'{e}rot while the \expres\ RVs have a median error of 38 \cms\ and are calibrated using a laser frequency comb.  The \expres\ observations are more stable, have smaller RV uncertainties, and return a best-fit RM model more aligned with literature values than previous attempts at measuring \cnce's stellar spin-orbit alignment.

With this robust measurement using \expres\ data, we can place constraints on the different proposed dynamical histories for the 55~Cnc system.  Studies have suggested that due to \cnc's distant stellar companion, one would expect all the planets around \cnc\ to precess as a rigid plane around the host star, resulting in all planets sharing a common planetary orbit axis that is misaligned with respect to the host star's spin axis \citep{kaib2011}.  Other simulations propose that secular resonances between \cnce\ and the other planets in the system led to excitation of \cnce's eccentricity and inclination, causing \cnce's orbital axis to be misaligned with the other planets' orbital axes as well as the host star's spin axis \citep{hanson2015}.  Analysis using the system's precession frequencies \citep{boue2014}, on the other hand, finds that one would expect to find \cnce's orbital axis close to aligned with its host star's spin axis \citep{boue2014_55cnc}, which is what our measurement shows.

Various theories for the migration pathways of USPs differ in terms of the predicted spin-orbit alignments.  Secular planet-planet interactions have been shown to give rise to tidal dissipation inside the planets and can operate under both a high-eccentricity and low-eccentricity mode.  The high-eccentricity mode leads to excitation of planetary eccentricities and inclinations through secular chaos, resulting in generally misaligned systems with large stellar spin-orbit alignment angles and USPs whose orbits have decayed to close-in, circular orbits \citep{petrovich2019}.  Planet-planet interactions can also lead to the inward migration of planets even in the case of modest eccentricities \citep{pu2019}.  This mechanism is less chaotic and likely results in USP orbits that are aligned to within a few tens of degrees.

Planetary obliquity tides have also been proposed as a source for tidal dissipation of USPs \citep{millholland2020}.  Here, planetary obliquity is used to refer to the angle between the planet's spin axis and the planet's orbital plane (as opposed to the stellar obliquity).  Planetary-obliquity-driven tidal dissipation has no requirements on initial planet eccentricities.  This mechanism predicts that a planet's orbit normal will become closer to aligned with the host star's spin axis as it migrates inwards, but may be misaligned with the orbit normals of other planets in the system.

The close alignment of \cnce's orbit normal with its host star's stellar axis preliminarily favors the low eccentricity \citep{pu2019} and planetary obliquity \citep{millholland2020} tide models.  It is of particular interest that \cnce\ is the only planet of the five known planets around \cnc\ that transits.  This suggests that though \cnce's orbit normal is aligned with the spin axis of its host star, it is possibly misaligned from those of the other planets.  This is expected from the planetary obliquity tides model if there existed some primordial misalignment of the stellar obliquity, for example due to \cnc's distant binary companion \citep{kaib2011}.

In the scenario with an initial misalignment, \cnce\ may have formed out of the protoplanetary disk with an orbital plane aligned with the outer planets, but as it migrated inward, \cnce\ became more influenced by the gravitational quadrupole moment of the star, causing it to become more aligned with the stellar equatorial plane as it settled into its final, short-period orbit \citep{millholland2020}.  The final state of \cnce\ would therefore be misaligned with the orbital planes of the outer, non-transiting planets, but aligned with the host star's spin equatorial plane as measured.

This type of system architecture for USPs is supported by previous studies.  \kepler\ USPs in multi-planet systems have been found to have larger mutual inclinations with the other outer planets \citep{dai2018}.  Recent analysis of HD~3167~b, an ultra-short period Super-Earth ($P = 0.959641 \substack{+1.1E-5\\ -1.2E-5}$ ;$M_\mathrm{p} = 5.02\pm0.38$\mearth), and HD~3167~c, a Sub-Neptune on a $~30$ day period, returned a similar system architecture \citep{christiansen2017,bourrier2021}.  The inner USP, HD~3167~b, was found to have an orbit normal closely aligned with its host star's spin axis ($\psi=29.5\substack{+7.2\\-9.4}^{\circ}$) while the further out planet, HD~3167~c, has an orbit nearly perpendicular to the orbit of HD~3167~b ($\psi=107.7\substack{+5.1\\-4.9}^{\circ}$).  We find that 55~Cnc similarly has a USP, \cnce, with an orbit aligned with the stellar equatorial plane though the alignment of the outer planets are unclear since they do not transit.

The close alignment of the ultra-short period, Super-Earth \cnce's orbit normal with its host star's spin axis places constraints on theories for how USPs migrate to their present day positions and how they interact with other planets in compact multi-planet systems.  This measurement additionally gives clues as to why none of the other known planets around \cnc\ transit and the possible role of \cnc's distant stellar companion.  This small-amplitude RM measurement of a Super-Earth is only now achievable because of new, ultra-stabilized spectrographs capable of delivering sub-meter-per-second precision.  More spin-orbit alignment measurements of USPs will help us to understand this unique population of planets and the nature of the planetary tidal dissipation that cause them to migrate so close-in to their host stars.

\section{Methods} \label{sec:methods}

\renewcommand{\figurename}{Supplementary Figure}
\setcounter{figure}{0}
\renewcommand{\tablename}{Supplementary Table}
\setcounter{table}{0}

\subsection{Photometric Data and Transit Fit}
\tess\ observed \cnc\ at a two-minute cadence during Sector 21 (Jan.\  21 - Feb.\ 18 2020) and a 20-s cadence during Sector 44 (Oct.\ 12 - Nov.\ 5 2021) and Sector 46 (Dec.\ 3-30 2021).  We use the publicly available light curves that have been detrended of common instrumental systematics \citep{jenkins2016}.  We found evidence of high-frequency correlated noise throughout the first sector (Sector 21) of observations, and therefore use only the later two sectors (i.e., Sectors 44 and 46) with 20-s exposure times.  In total, we used 199,838 \tess\ cadences of \cnc.

Both sectors of \tess\ data are fit for simultaneously with a Hamiltonian MCMC implemented using \exoplanet, \pymc, and \theano\ \citep{exoplanet, pymc3, theano}.  The fit is parameterized using the stellar mass and radius rather than just the stellar density since the planet has a low eccentricity and there exists direct interferometric constraints on the stellar radius of 55~Cnc \citep{vonbraun2011}.  The Gaussian priors for the stellar mass and radius are given in Supplementary~Table~1.  We additionally fit for the orbital period, $P$, transit mid-time, $T_0$, planet-to-star radius ratio, $R_\mathrm{p}/R_\star$, and impact parameter, $b$. Each sector is also fit to an independent GP model using a Matérn-3/2 kernel parametrized by an amplitude $\sigma_\mathrm{GP}$ and a timescale  $\tau_\mathrm{GP}$ to account for any background variations in the light curve \citep{celerite1, celerite2}. The model also includes a mean flux and a white noise term to inflate the photon noise uncertainties from \tess{}.

The model included quadratic limb darkening, with coefficients determined using the \ldtk\ code package \citep{parviainen2015}. This code uses the \phoenix\ stellar atmosphere models \citep{husser2013} to generate observed disk intensities based on the \tess\ bandpass and input stellar parameters $T_\mathrm{eff}$, Fe/H, and $\log{g}$ that were obtained from spectroscopic analysis (Supplementary~Table~1).  The limb-darkening coefficients were fit for as part of the model and were constrained using Gaussian priors, though the uncertainties were inflated by a factor of five to account for the uncertainties in the stellar model. We ran two independent chains for 1500 tuning steps and 1000 production steps.  We verified that convergence was reached by making sure the $\hat{R}$ metric, as defined in \cite{gelman2003}, was below a conservative 1.001 threshold.  We additionally ensured that the effective number of samples per parameter was ${>}200$. The \tess\ data and best-fit transit model is shown in Supplementary~Figure~1.  The resultant transit fit parameters, which agree with literature values \citep{bourrier2018,demory2016,kipping2020}, are given in Table~\ref{tab:transitFit} with the GP parameters given in Table~\ref{tab:fit_nuisance_params}.

\begin{figure}[htp]
\centering
\includegraphics[width=0.8\textwidth]{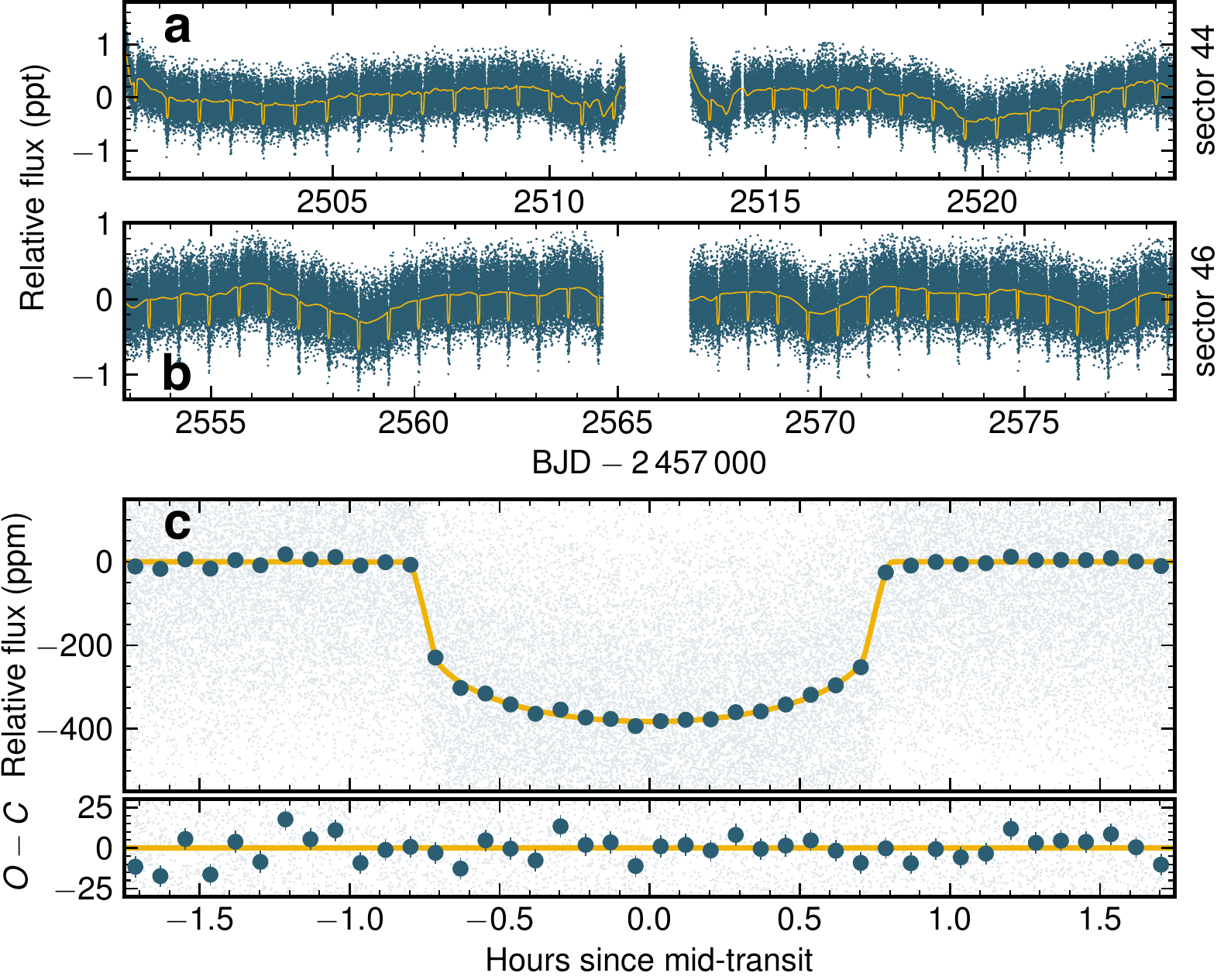}
\centering
\caption{Fit to two sectors of \tess\ photometry.  Panels a and b show the full time series of 20-second exposures in blue with the best-fit transit model over plotted in yellow.  Panel c shows the phase-folded light curve with the GP model accounting for background variations removed.  In panel c, each individual exposure is shown as a light blue point.  Dark blue points are phase binned into 5-minute averages and are shown as a visual aid.  Phase binned points are calculated as the average of all data points within the phase bin weighted by their errors, $\sigma_i$, with associated errors given by $\sqrt{\frac{1}{\sum 1/\sigma^2_i}}$.  Phase-binned residuals to the transit model are shown in the bottom most panel.
}
\label{fig:phot_data}
\end{figure}

\subsection{Spectroscopic Data}
\expres\ observed \cnce\ during transit on Jan.\ 31, 2022 (28 observations) and Apr.\ 6, 2022 (27 observations).  \expres\ is a stabilized, next-generation, optical ($390-780$ nm) spectrograph with a high median resolving power of $R\sim137,000$ \citep{jurgenson2016} mounted on the 4.3-m Lowell Discovery Telescope (\ldt) \citep{levine2012}.  It has a measured instrument stability of 3-7 \cms\ and returns an intra-night scatter of 10-40 \cms\ on-sky for select quiet stars \citep{blackman2020, zhao2022}.

\expres\ observations were terminated after they reached an SNR of 225, which corresponds to a resultant analytical RV error of approximately 35 \cms\ \citep{petersburg2020}.  Most observations took between 4 and 9 minutes.  The last five observations on Apr.\ 6, which fall after the transit, required longer exposure times of 15-17 minutes to reach the target SNR as the weather quickly degraded.  RVs are derived using a template-matching, chunk-by-chunk technique.  These RV measurements are given in Supplementary~Table~2.

\subsection{\cnc\ Stellar Properties} \label{sec:star}
We also use the \expres\ spectra to derive stellar properties of \cnc\ \citep{brewer2016}.  The analysis uses the forward modeling tool, Spectroscpoy Made Easy (SME) \citep{piskunov2017}, to fit both the global stellar properties and individual abundances of 15 elements using more than 7000 atomic and molecular lines over $\sim$350 \AA\ of spectrum.  The method first fits for \teff, \logg, rotational broadening, and a scaled solar abundance pattern [M/H], with only calcium, silicon, and titanium allowed to vary independently.  The temperature of this model is perturbed $\pm 100$ K, and the parameters are re-derived.  The global parameters are then fixed to those found while the abundances of 15 elements (C, N, O, Na, Mg, Al, Si, Ca, Ti, V, Cr, Mn, Fe, Ni, and Y) are allowed to vary.  These steps are then repeated, this time starting with the newly-derived abundance pattern.  Finally, a temperature and gravity dependent macroturbulence is assigned and we solve for \vsini\ with all other parameters fixed.

In total, 290 \expres\ \cnc\ spectra were analyzed in this way; the $\chi^2$ weighted averages adopted for the parameters can be found in Supplementary~Table~1.  The distribution of returned model parameters gives us a measure of the model fit for this star but does not take into account uncertainties in the model itself.  The standard deviations in returned \teff, \logg, and \vsini\ were 5 K, 0.015 dex, and 0.05 km/s respectively. We therefore adopt the larger uncertainties that take into account model uncertainties across a range of stellar parameters \citep{brewer2016}.

\begin{table}[h]
\scriptsize
\caption{\cnc\ Stellar Parameters Derived using \sme}
\label{tab:star}
\begin{center}
\begin{tabular}{l c c l l}
\hline
\hline
Parameter & Symbol & Value & Unit & Reverence \\
\hline
Radius & $R_{\star}$ & 0.943$\pm$0.010 & \rsun & von Braun et al.\ 2011 \\
Mass & $M_{\star}$ & 0.905$\pm$0.015 & \msun & von Braun et al.\ 2011 \\
Rotation Period & $P_{\star}$ & 38.8$\pm$0.05 & days & Bourrier et al.\ 2018 \\
\hline
\textbf{This Work} & & & \\
Projected rotational velocity & \vsini & 1.94$\pm$0.50\ & \kms \\
Effective temperature & \teff & 5272$\pm$24\ & K \\
Surface gravity & \logg & 4.38$\pm$0.05\ & m s\textsuperscript{-2} \\
Metallicity & {[}Fe/H{]} & 0.37$\pm$0.02\ & \\
\hline
\end{tabular}
\end{center}
\end{table}

\begin{deluxetable}{ccc}[b!]
\tabletypesize{\scriptsize}
\tablecaption{Chunk-by-Chunk RVs \label{tab:cbc_rvs}}
\tablehead{
\colhead{Time [BMJD\textsubscript{UTC}]} & \colhead{RV [\ms]} & \colhead{Err. [\ms]}
}
\startdata
59610.371 & 87.665 & 0.345 \\ 
59610.376 & 86.737 & 0.337 \\ 
59610.381 & 86.493 & 0.3 \\ 
 \vdots & \vdots & \vdots \\
\enddata
\tablecomments{A stub of this table is provided here for reference; the full RV data set is available online.}
\end{deluxetable}

\subsection{Rossiter-Mclaughlin Fit}
We model the Rossiter-McLaughlin signal in the \expres\ radial velocity data using \ellc\ \citep{ohta2005, maxted2016}. The \emcee\ ensemble sampler is used to perform MCMC sampling to find the range of parameters compatible with the data \citep{emcee}. In the MCMC sampling, we allow the projected stellar rotation, \vsini, and sky-projected stellar obliquity, $\lambda$, to vary uniformly in the range 0-20 \kms\ and -180 to 180 degrees, respectively.  By using uniform priors for \vsini\ rather than priors based on the value determined by SME, we are able to get independent measures of \vsini\ from both SME and the RM fit.

For each transit, our model includes a white noise term to inflate the photon noise uncertainty as well as a linear model with parameters $y_0$ (intercept) and $m$ (slope) to account for the stellar reflex motion due to the orbiting planets. The slope is a free parameter that is fit to the out-of-transit data at every step in the MCMC. The intercept is then determined from a weighted average of the out-of-transit residuals after subtracting the slope.

The transit also gives rise to a second radial velocity anomaly due to turbulent convective motion on the surface of the star \citep{cegla2016}. We include a model for the convective blueshift variation \citep{shporer2011} and assume a net convective blueshift centered on $V_\mathrm{CB} =-150$ \ms \citep{meunier2017, liebing2021} with a standard deviation of 100 \ms. The contribution of the convective blueshift model has a semi-amplitude of 1.3 \cms\ and is shown as a gray dotted line in Figure~1.

The orbital period $P$, transit time $t_0$, scaled separation $a/R_\star$, radius ratio $r/R_\star$, and orbital inclination $i$ are updated in every step of the MCMC by sampling from a multivariate normal distribution, the mean and covariance properties of which are determined from the posterior distribution obtained from the \tess\ analysis. This effectively places Gaussian priors on the transit parameters, but also takes into account their covariance.

We ran an ensemble of 300 \emcee\ ``walkers'' for 2000 steps each, discarded 750 samples as a burn-in, thinned the chains by the autocorrelation length, and finally merged the chains together to obtain 3300 independent samples. The median values and $68\%$ confidence interval for the fit for RM parameters are reported in Table~\ref{tab:transitFit}.  The range of models compatible with the data to the 99\textsuperscript{th} percentile are illustrated in Figure~1 by the yellow shading. 

The Rossiter-McLaughlin effect is only able to measure the sky-projection of the stellar obliquity, $\lambda$.  The 3D obliquity, $\psi$ can be recovered using an independent measurement of the inclination of the star, $i_\star$. We derive an estimate of $i_\star = 75\substack{+11\\ -17}^\circ$ by combining measurements of the stellar radius $R_\star$, stellar rotation period $P_\mathrm{rot}$, and the projected rotation \vsini{} from the RM fit \citep{masuda2020}. The distribution of $i_\star$ is broad, but there is a preference for a somewhat inclined star, where $i_\star = 90^\circ$ means the spin angular momentum vector is perpendicular to our line of sight. The combination of $i_\star$, $\lambda$, and orbital inclination $i$ gives a de-projected stellar obliquity of $\psi = 23\substack{+15\\ -12}^\circ$, which is consistent with a prograde, low stellar obliquity orbit for \cnce{}.


Our detection of a Rossiter-McLaughlin effect is statistically supported.  Our best fit value of \vsini\ over its corresponding error is 4.3$\sigma$.  We implement leave-one-out cross-validation to compare the RM model to the null hypothesis of no detection, i.e. a straight line \citep{andrae2010}. This returns a $\chi^2$ of 70.3 for an RM model vs. 86.9 for a model without RM, implying that the RM model better predicts the ``unseen'' left out data and is therefore the better choice of model.  The Bayesian information criterion based on the maximum likelihood fit to the full data set also favors the RM model over a straight line by 8.5, where a difference greater than 6 is typically considered significant \citep{schwarz1978}.

We use Gaussian process (GP) regression bootstrapping to establish a false alarm probability for our measured RM effect.  We generate 1000 random pairs of data samples using a GP with a Matérn-3/2 covariance function whose hyperparameters are set to reflect the noise properties of the observed \expres\ data. These data have no injected RM signal as the samples have been drawn from the prior distribution over datasets. Of these realizations, only 0.3\% return a best fit RM model with a solution consistent to the RM fit with the real data, illustrated by the yellow contour in Supplementary~Figure~2.  Our detection therefore has a false alarm probability of 0.3\%.

\begin{figure}[h!]
\centering
\includegraphics[width=0.375\textwidth]{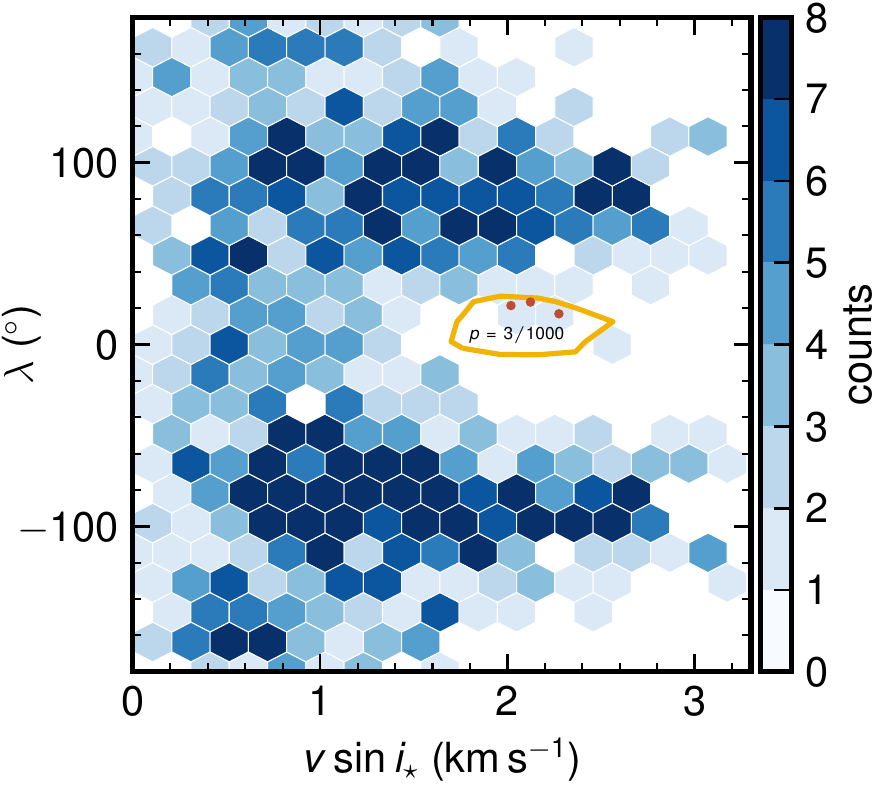}
\caption{A 2D histogram of the results from maximum likelihood RM fits to 1000 data samples using Gaussian process regression bootstrapping. The yellow contour represents the 68\% credible interval of the fit to the original data. Three out of 1000 versions of the generated mock data (red dots) show signals consistent with our original data, signifying a 0.3\% false positive probability. One point is outside the range of the figure at $v\sin{i_\star} = 4.09$ \kms.
}
\label{fig:bootstrapping}
\end{figure}

\section{Data Availability}
The \expres\ radial velocities used in this study are published as part of the supplementary information.  The \tess\ data used in this study is publicly available and can be obtained from the Mikulski Archive for Space Telescopes (MAST; https://archive.stsci.edu/missions-and-data/tess).

\section{Code Availability}
The code associated with this work used only open source software.  This research made use of \code{SciPy} \citep{scipy}, \code{NumPy} \citep{numpy, numpy2}, \code{Astropy} \citep{astropy2013,astropy2018}, \lightkurve\ \citep{lightkurve}, \starry\ \citep{luger2018}, \emcee\ \citep{emcee}, \celerite\ \citep{celerite1,celerite2}, \ellc\ \citep{maxted2016}, \ldtk\ \citep{parviainen2015}.  This research also made use of \textsf{exoplanet} \citep{exoplanet} and its
dependencies \citep{agol2020, astropy2013, astropy2018, kipping2013, luger2018, pymc3, theano}.

\section{Acknowledgements}
We thank the anonymous referees for improving this manuscript with their thoughtful feedback.  
These results made use of data provided by the \expres\ team using the EXtreme PREcision Spectrograph at the Lowell Discovery telescope, Lowell Observatory. Lowell is a private, non-profit institution dedicated to astrophysical research and public appreciation of astronomy and operates the LDT in partnership with Boston University, the University of Maryland, the University of Toledo, Northern Arizona University and Yale University. \expres\ was designed and built at Yale with financial support from NSF MRI-1429365, NSF ATI-1509436 and Yale University. Research with \expres\ is possible thanks to the generous support from NSF AST-2009528, NSF 1616086, NASA 80NSSC18K0443, the Heising-Simons Foundation, and an anonymous donor in the Yale alumni community.    
This paper includes data collected with the TESS mission, obtained from the MAST data archive at the Space Telescope Science Institute (STScI). Funding for the TESS mission is provided by the NASA Explorer Program. STScI is operated by the Association of Universities for Research in Astronomy, Inc., under NASA contract NAS 5–26555. 
V.K. and J.L. acknowledge support from NSF awards AST-2009501 and AST-2009343.  
J.M.B. acknowledges support from NASA grants 80NSSC21K0009 and  80NSSC21K0571.  
R.M.R. acknowledges support from the Heising-Simons 51 Pegasi b Postdoctoral Fellowship. 

\section{Author Contributions}
L.L.Z. designed the project and drafted the manuscript.  V.K. led the analysis and drafted the manuscript.  S.C.M. contributed to the scientific interpretation.  C.H. processed the \tess\ data.  L.L.Z., J.M.B., J.L., A.E.S., R.M.R., S.H.C.C., S.A.W., D.A.F. are members of the \expres\ team that built and commissioned \expres, maintain the instrument for high precision work, and supervise the data reduction pipeline.  D.A.F. is the PI of the \expres\ Team and derived the \expres\ RVs.  J.M.B. ran the stellar parameter analysis.  L.L.Z., V.K., J.M.B.,  J.L.,  S.H.C.C.,  S.A.W., and D.A.F. contributed to the \expres\ observations.

\section{Competing Interests}
The authors declare no competing interests.

\bibliographystyle{naturemag} 
\bibliography{main} 

\begin{thebibliography}{}
\expandafter\ifx\csname natexlab\endcsname\relax\def\natexlab#1{#1}\fi
\providecommand{\url}[1]{\href{#1}{#1}}
\providecommand{\dodoi}[1]{doi:~\href{http://doi.org/#1}{\nolinkurl{#1}}}
\providecommand{\doeprint}[1]{\href{http://ascl.net/#1}{\nolinkurl{http://ascl.net/#1}}}
\providecommand{\doarXiv}[1]{\href{https://arxiv.org/abs/#1}{\nolinkurl{https://arxiv.org/abs/#1}}}

\bibitem[{{Agol} {et~al.}(2020){Agol}, {Luger}, \& {Foreman-Mackey}}]{agol2020}
{Agol}, E., {Luger}, R., \& {Foreman-Mackey}, D. 2020, \aj, 159, 123,
  \dodoi{10.3847/1538-3881/ab4fee}

\bibitem[{{Albrecht} {et~al.}(2022){Albrecht}, {Dawson}, \&
  {Winn}}]{albrecht2022}
{Albrecht}, S.~H., {Dawson}, R.~I., \& {Winn}, J.~N. 2022, arXiv e-prints,
  arXiv:2203.05460.
\newblock \doarXiv{2203.05460}

\bibitem[{Andrae {et~al.}(2010)Andrae, {Schulze-Hartung}, \&
  Melchior}]{andrae2010}
Andrae, R., {Schulze-Hartung}, T., \& Melchior, P. 2010, arXiv:1012.3754
  [astro-ph, physics:physics, stat].
\newblock \doarXiv{1012.3754}

\bibitem[{{Astropy Collaboration} {et~al.}(2013){Astropy Collaboration},
  {Robitaille}, {Tollerud}, {Greenfield}, {Droettboom}, {Bray}, {Aldcroft},
  {Davis}, {Ginsburg}, {Price-Whelan}, {Kerzendorf}, {Conley}, {Crighton},
  {Barbary}, {Muna}, {Ferguson}, {Grollier}, {Parikh}, {Nair}, {Unther},
  {Deil}, {Woillez}, {Conseil}, {Kramer}, {Turner}, {Singer}, {Fox}, {Weaver},
  {Zabalza}, {Edwards}, {Azalee Bostroem}, {Burke}, {Casey}, {Crawford},
  {Dencheva}, {Ely}, {Jenness}, {Labrie}, {Lim}, {Pierfederici}, {Pontzen},
  {Ptak}, {Refsdal}, {Servillat}, \& {Streicher}}]{astropy2013}
{Astropy Collaboration}, {Robitaille}, T.~P., {Tollerud}, E.~J., {et~al.} 2013,
  \aap, 558, A33, \dodoi{10.1051/0004-6361/201322068}

\bibitem[{{Blackman} {et~al.}(2020){Blackman}, {Fischer}, {Jurgenson},
  {Sawyer}, {McCracken}, {Szymkowiak}, {Petersburg}, {Ong}, {Brewer}, {Zhao},
  {Leet}, {Buchhave}, {Tronsgaard}, {Llama}, {Sawyer}, {Davis}, {Cabot},
  {Shao}, {Trahan}, {Nemati}, {Genoni}, {Pariani}, {Riva}, {Fournier}, \&
  {Pawluczyk}}]{blackman2020}
{Blackman}, R.~T., {Fischer}, D.~A., {Jurgenson}, C.~A., {et~al.} 2020, \aj,
  159, 238, \dodoi{10.3847/1538-3881/ab811d}

\bibitem[{{Bou{\'e}} \& {Fabrycky}(2014{\natexlab{a}})}]{boue2014}
{Bou{\'e}}, G., \& {Fabrycky}, D.~C. 2014{\natexlab{a}}, \apj, 789, 111,
  \dodoi{10.1088/0004-637X/789/2/111}

\bibitem[{{Bou{\'e}} \& {Fabrycky}(2014{\natexlab{b}})}]{boue2014_55cnc}
{Bou{\'e}}, G., \& {Fabrycky}, D.~C. 2014{\natexlab{b}}, in Complex Planetary
  Systems, Proceedings of the International Astronomical Union, Vol. 310,
  62--65

\bibitem[{{Bourrier} \& {H{\'e}brard}(2014)}]{bourrier2014}
{Bourrier}, V., \& {H{\'e}brard}, G. 2014, \aap, 569, A65,
  \dodoi{10.1051/0004-6361/201424266}

\bibitem[{{Bourrier} {et~al.}(2018){Bourrier}, {Dumusque}, {Dorn}, {Henry},
  {Astudillo-Defru}, {Rey}, {Benneke}, {H{\'e}brard}, {Lovis}, {Demory},
  {Moutou}, \& {Ehrenreich}}]{bourrier2018}
{Bourrier}, V., {Dumusque}, X., {Dorn}, C., {et~al.} 2018, \aap, 619, A1,
  \dodoi{10.1051/0004-6361/201833154}

\bibitem[{{Bourrier} {et~al.}(2021){Bourrier}, {Lovis}, {Cretignier}, {Allart},
  {Dumusque}, {Delisle}, {Deline}, {Sousa}, {Adibekyan}, {Alibert}, {Barros},
  {Borsa}, {Cristiani}, {Demangeon}, {Ehrenreich}, {Figueira}, {Gonz{\'a}lez
  Hern{\'a}ndez}, {Lendl}, {Lillo-Box}, {Lo Curto}, {Di Marcantonio},
  {Martins}, {M{\'e}gevand}, {Mehner}, {Micela}, {Molaro}, {Oshagh}, {Palle},
  {Pepe}, {Poretti}, {Rebolo}, {Santos}, {Scandariato}, {Seidel}, {Sozzetti},
  {Su{\'a}rez Mascare{\~n}o}, \& {Zapatero Osorio}}]{bourrier2021}
{Bourrier}, V., {Lovis}, C., {Cretignier}, M., {et~al.} 2021, \aap, 654, A152,
  \dodoi{10.1051/0004-6361/202141527}

\bibitem[{{Brewer} {et~al.}(2016){Brewer}, {Fischer}, {Valenti}, \&
  {Piskunov}}]{brewer2016}
{Brewer}, J.~M., {Fischer}, D.~A., {Valenti}, J.~A., \& {Piskunov}, N. 2016,
  \apjs, 225, 32, \dodoi{10.3847/0067-0049/225/2/32}

\bibitem[{{Butler} {et~al.}(1997){Butler}, {Marcy}, {Williams}, {Hauser}, \&
  {Shirts}}]{butler1997}
{Butler}, R.~P., {Marcy}, G.~W., {Williams}, E., {Hauser}, H., \& {Shirts}, P.
  1997, \apjl, 474, L115, \dodoi{10.1086/310444}

\bibitem[{{Cegla} {et~al.}(2016){Cegla}, {Oshagh}, {Watson}, {Figueira},
  {Santos}, \& {Shelyag}}]{cegla2016}
{Cegla}, H.~M., {Oshagh}, M., {Watson}, C.~A., {et~al.} 2016, \apj, 819, 67,
  \dodoi{10.3847/0004-637X/819/1/67}

\bibitem[{{Christiansen} {et~al.}(2017){Christiansen}, {Vanderburg}, {Burt},
  {Fulton}, {Batygin}, {Benneke}, {Brewer}, {Charbonneau}, {Ciardi}, {Collier
  Cameron}, {Coughlin}, {Crossfield}, {Dressing}, {Greene}, {Howard}, {Latham},
  {Molinari}, {Mortier}, {Mullally}, {Pepe}, {Rice}, {Sinukoff}, {Sozzetti},
  {Thompson}, {Udry}, {Vogt}, {Barman}, {Batalha}, {Bouchy}, {Buchhave},
  {Butler}, {Cosentino}, {Dupuy}, {Ehrenreich}, {Fiorenzano}, {Hansen},
  {Henning}, {Hirsch}, {Holden}, {Isaacson}, {Johnson}, {Knutson}, {Kosiarek},
  {L{\'o}pez-Morales}, {Lovis}, {Malavolta}, {Mayor}, {Micela}, {Motalebi},
  {Petigura}, {Phillips}, {Piotto}, {Rogers}, {Sasselov}, {Schlieder},
  {S{\'e}gransan}, {Watson}, \& {Weiss}}]{christiansen2017}
{Christiansen}, J.~L., {Vanderburg}, A., {Burt}, J., {et~al.} 2017, \aj, 154,
  122, \dodoi{10.3847/1538-3881/aa832d}

\bibitem[{{Dai} {et~al.}(2018){Dai}, {Masuda}, \& {Winn}}]{dai2018}
{Dai}, F., {Masuda}, K., \& {Winn}, J.~N. 2018, \apjl, 864, L38,
  \dodoi{10.3847/2041-8213/aadd4f}

\bibitem[{{Dawson} \& {Fabrycky}(2010)}]{dawson2010}
{Dawson}, R.~I., \& {Fabrycky}, D.~C. 2010, \apj, 722, 937,
  \dodoi{10.1088/0004-637X/722/1/937}

\bibitem[{{Demory} {et~al.}(2011){Demory}, {Gillon}, {Deming}, \&
  {Seager}}]{demory2011}
{Demory}, B.-O., {Gillon}, M., {Deming}, D., \& {Seager}, S. 2011, in
  AAS/Division for Extreme Solar Systems Abstracts, Vol.~2, AAS/Division for
  Extreme Solar Systems Abstracts, 17.04

\bibitem[{Demory {et~al.}(2016)Demory, Gillon, {de Wit}, Madhusudhan, Bolmont,
  Heng, Kataria, Lewis, Hu, Krick, Stamenkovi{\'c}, Benneke, Kane, \&
  Queloz}]{demory2016}
Demory, B.-O., Gillon, M., {de Wit}, J., {et~al.} 2016, Nature, 532, 207,
  \dodoi{10.1038/nature17169}

\bibitem[{{Fischer} {et~al.}(2008){Fischer}, {Marcy}, {Butler}, {Vogt},
  {Laughlin}, {Henry}, {Abouav}, {Peek}, {Wright}, {Johnson}, {McCarthy}, \&
  {Isaacson}}]{fischer2008}
{Fischer}, D.~A., {Marcy}, G.~W., {Butler}, R.~P., {et~al.} 2008, \apj, 675,
  790, \dodoi{10.1086/525512}

\bibitem[{{Foreman-Mackey}(2018)}]{celerite2}
{Foreman-Mackey}, D. 2018, Research Notes of the American Astronomical Society,
  2, 31, \dodoi{10.3847/2515-5172/aaaf6c}

\bibitem[{{Foreman-Mackey} {et~al.}(2017){Foreman-Mackey}, {Agol},
  {Ambikasaran}, \& {Angus}}]{celerite1}
{Foreman-Mackey}, D., {Agol}, E., {Ambikasaran}, S., \& {Angus}, R. 2017, \aj,
  154, 220, \dodoi{10.3847/1538-3881/aa9332}

\bibitem[{{Foreman-Mackey} {et~al.}(2013){Foreman-Mackey}, {Hogg}, {Lang}, \&
  {Goodman}}]{emcee}
{Foreman-Mackey}, D., {Hogg}, D.~W., {Lang}, D., \& {Goodman}, J. 2013, \pasp,
  125, 306, \dodoi{10.1086/670067}

\bibitem[{Foreman-Mackey {et~al.}(2020)Foreman-Mackey, Luger, Czekala, Agol,
  Price-Whelan, Brandt, Barclay, \& Bouma}]{exoplanet}
Foreman-Mackey, D., Luger, R., Czekala, I., {et~al.} 2020,
  exoplanet-dev/exoplanet v0.4.0, \dodoi{10.5281/zenodo.1998447}.
\newblock \url{https://doi.org/10.5281/zenodo.1998447}

\bibitem[{Gelman {et~al.}(2003)Gelman, Carlin, Stern, \& Rubin}]{gelman2003}
Gelman, A., Carlin, J., Stern, H., \& Rubin, D. 2003, Bayesian Data Analysis,
  Second Edition, Chapman \& Hall/CRC Texts in Statistical Science (Taylor \&
  Francis).
\newblock \url{https://books.google.co.uk/books?id=TNYhnkXQSjAC}

\bibitem[{{Hansen} \& {Zink}(2015)}]{hanson2015}
{Hansen}, B. M.~S., \& {Zink}, J. 2015, \mnras, 450, 4505,
  \dodoi{10.1093/mnras/stv916}

\bibitem[{Husser {et~al.}(2013)Husser, {Wende-von Berg}, Dreizler, Homeier,
  Reiners, Barman, \& Hauschildt}]{husser2013}
Husser, T.-O., {Wende-von Berg}, S., Dreizler, S., {et~al.} 2013, Astronomy and
  Astrophysics, 553, A6, \dodoi{10.1051/0004-6361/201219058}

\bibitem[{{Jenkins}(2016)}]{jenkins2016}
{Jenkins}, J.~M. 2016, IAU Focus Meeting, 29A, 210,
  \dodoi{10.1017/S1743921316002842}

\bibitem[{{Jurgenson} {et~al.}(2016){Jurgenson}, {Fischer}, {McCracken},
  {Sawyer}, {Szymkowiak}, {Davis}, {Muller}, \& {Santoro}}]{jurgenson2016}
{Jurgenson}, C., {Fischer}, D., {McCracken}, T., {et~al.} 2016, in \procspie,
  Vol. 9908, Ground-based and Airborne Instrumentation for Astronomy VI, 99086T

\bibitem[{{Kaib} {et~al.}(2011){Kaib}, {Raymond}, \& {Duncan}}]{kaib2011}
{Kaib}, N.~A., {Raymond}, S.~N., \& {Duncan}, M.~J. 2011, \apjl, 742, L24,
  \dodoi{10.1088/2041-8205/742/2/L24}

\bibitem[{{Kipping} \& {Jansen}(2020)}]{kipping2020}
{Kipping}, D., \& {Jansen}, T. 2020, Research Notes of the American
  Astronomical Society, 4, 170, \dodoi{10.3847/2515-5172/abbc0f}

\bibitem[{{Kipping}(2013)}]{kipping2013}
{Kipping}, D.~M. 2013, \mnras, 435, 2152, \dodoi{10.1093/mnras/stt1435}

\bibitem[{Kunovac~Hod{\v z}i{\'c} {et~al.}(2021)Kunovac~Hod{\v z}i{\'c},
  Triaud, Cegla, Chaplin, \& Davies}]{kunovachodzic2021}
Kunovac~Hod{\v z}i{\'c}, V., Triaud, A. H. M.~J., Cegla, H.~M., Chaplin, W.~J.,
  \& Davies, G.~R. 2021, Monthly Notices of the Royal Astronomical Society,
  502, 2893, \dodoi{10.1093/mnras/stab237}

\bibitem[{{Levine} {et~al.}(2012){Levine}, {Bida}, {Chylek}, {Collins},
  {DeGroff}, {Dunham}, {Lotz}, {Venetiou}, \& {Zoonemat Kermani}}]{levine2012}
{Levine}, S.~E., {Bida}, T.~A., {Chylek}, T., {et~al.} 2012, in Society of
  Photo-Optical Instrumentation Engineers (SPIE) Conference Series, Vol. 8444,
  Ground-based and Airborne Telescopes IV, 844419

\bibitem[{Liebing {et~al.}(2021)Liebing, Jeffers, Reiners, \&
  Zechmeister}]{liebing2021}
Liebing, F., Jeffers, S.~V., Reiners, A., \& Zechmeister, M. 2021, Astronomy
  and Astrophysics, 654, A168, \dodoi{10.1051/0004-6361/202039607}

\bibitem[{{Lightkurve Collaboration} {et~al.}(2018){Lightkurve Collaboration},
  {Cardoso}, {Hedges}, {Gully-Santiago}, {Saunders}, {Cody}, {Barclay}, {Hall},
  {Sagear}, {Turtelboom}, {Zhang}, {Tzanidakis}, {Mighell}, {Coughlin}, {Bell},
  {Berta-Thompson}, {Williams}, {Dotson}, \& {Barentsen}}]{lightkurve}
{Lightkurve Collaboration}, {Cardoso}, J.~V.~d.~M., {Hedges}, C., {et~al.}
  2018, {Lightkurve: Kepler and TESS time series analysis in Python},
  Astrophysics Source Code Library.
\newblock \doeprint{1812.013}

\bibitem[{{L{\'o}pez-Morales} {et~al.}(2014){L{\'o}pez-Morales}, {Triaud},
  {Rodler}, {Dumusque}, {Buchhave}, {Harutyunyan}, {Hoyer}, {Alonso}, {Gillon},
  {Kaib}, {Latham}, {Lovis}, {Pepe}, {Queloz}, {Raymond}, {S{\'e}gransan},
  {Waldmann}, \& {Udry}}]{lopezmorales2014}
{L{\'o}pez-Morales}, M., {Triaud}, A. H.~M.~J., {Rodler}, F., {et~al.} 2014,
  \apjl, 792, L31, \dodoi{10.1088/2041-8205/792/2/L31}

\bibitem[{{Luger} {et~al.}(2019){Luger}, {Agol}, {Foreman-Mackey}, {Fleming},
  {Lustig-Yaeger}, \& {Deitrick}}]{luger2018}
{Luger}, R., {Agol}, E., {Foreman-Mackey}, D., {et~al.} 2019, \aj, 157, 64,
  \dodoi{10.3847/1538-3881/aae8e5}

\bibitem[{{Marcy} {et~al.}(2002){Marcy}, {Butler}, {Fischer}, {Laughlin},
  {Vogt}, {Henry}, \& {Pourbaix}}]{marcy2002}
{Marcy}, G.~W., {Butler}, R.~P., {Fischer}, D.~A., {et~al.} 2002, \apj, 581,
  1375, \dodoi{10.1086/344298}

\bibitem[{Masuda \& Winn(2020)}]{masuda2020}
Masuda, K., \& Winn, J.~N. 2020, The Astronomical Journal, 159, 81,
  \dodoi{10.3847/1538-3881/ab65be}

\bibitem[{Maxted(2016)}]{maxted2016}
Maxted, P. F.~L. 2016, Astronomy and Astrophysics, 591, A111,
  \dodoi{10.1051/0004-6361/201628579}

\bibitem[{{McArthur} {et~al.}(2004){McArthur}, {Endl}, {Cochran}, {Benedict},
  {Fischer}, {Marcy}, {Butler}, {Naef}, {Mayor}, {Queloz}, {Udry}, \&
  {Harrison}}]{mcarthur2004}
{McArthur}, B.~E., {Endl}, M., {Cochran}, W.~D., {et~al.} 2004, \apjl, 614,
  L81, \dodoi{10.1086/425561}

\bibitem[{{Meunier} {et~al.}(2017){Meunier}, {Mignon}, \&
  {Lagrange}}]{meunier2017}
{Meunier}, N., {Mignon}, L., \& {Lagrange}, A.~M. 2017, \aap, 607, A124,
  \dodoi{10.1051/0004-6361/201731017}

\bibitem[{{Millholland} \& {Spalding}(2020)}]{millholland2020}
{Millholland}, S.~C., \& {Spalding}, C. 2020, \apj, 905, 71,
  \dodoi{10.3847/1538-4357/abc4e5}

\bibitem[{Ohta {et~al.}(2005)Ohta, Taruya, \& Suto}]{ohta2005}
Ohta, Y., Taruya, A., \& Suto, Y. 2005, The Astrophysical Journal, 622, 1118,
  \dodoi{10.1086/428344}

\bibitem[{Oliphant(2006--)}]{numpy}
Oliphant, T. 2006--, {NumPy}: A guide to {NumPy}, USA: Trelgol Publishing.
\newblock \url{http://www.numpy.org/}

\bibitem[{{Parviainen} \& {Aigrain}(2015)}]{parviainen2015}
{Parviainen}, H., \& {Aigrain}, S. 2015, \mnras, 453, 3821,
  \dodoi{10.1093/mnras/stv1857}

\bibitem[{{Pepe} {et~al.}(2013){Pepe}, {Cristiani}, {Rebolo}, {Santos},
  {Dekker}, {M{\'e}gevand}, {Zerbi}, {Cabral}, {Molaro}, {Di Marcantonio},
  {Abreu}, {Affolter}, {Aliverti}, {Allende Prieto}, {Amate}, {Avila},
  {Baldini}, {Bristow}, {Broeg}, {Cirami}, {Coelho}, {Conconi}, {Coretti},
  {Cupani}, {D'Odorico}, {De Caprio}, {Delabre}, {Dorn}, {Figueira}, {Fragoso},
  {Galeotta}, {Genolet}, {Gomes}, {Gonz{\'a}lez Hern{\'a}ndez}, {Hughes},
  {Iwert}, {Kerber}, {Landoni}, {Lizon}, {Lovis}, {Maire}, {Mannetta},
  {Martins}, {Monteiro}, {Oliveira}, {Poretti}, {Rasilla}, {Riva}, {Santana
  Tschudi}, {Santos}, {Sosnowska}, {Sousa}, {Span{\`o}}, {Tenegi}, {Toso},
  {Vanzella}, {Viel}, \& {Zapatero Osorio}}]{pepe2013}
{Pepe}, F., {Cristiani}, S., {Rebolo}, R., {et~al.} 2013, The Messenger, 153, 6

\bibitem[{{Petersburg} {et~al.}(2020){Petersburg}, {Ong}, {Zhao}, {Blackman},
  {Brewer}, {Buchhave}, {Cabot}, {Davis}, {Jurgenson}, {Leet}, {McCracken},
  {Sawyer}, {Sharov}, {Tronsgaard}, {Szymkowiak}, \&
  {Fischer}}]{petersburg2020}
{Petersburg}, R.~R., {Ong}, J.~M.~J., {Zhao}, L.~L., {et~al.} 2020, \aj, 159,
  187, \dodoi{10.3847/1538-3881/ab7e31}

\bibitem[{{Petrovich} {et~al.}(2019){Petrovich}, {Deibert}, \&
  {Wu}}]{petrovich2019}
{Petrovich}, C., {Deibert}, E., \& {Wu}, Y. 2019, \aj, 157, 180,
  \dodoi{10.3847/1538-3881/ab0e0a}

\bibitem[{Piskunov \& Valenti(2017)}]{piskunov2017}
Piskunov, N., \& Valenti, J.~A. 2017, Astronomy and Astrophysics, 597, A16,
  \dodoi{10.1051/0004-6361/201629124}

\bibitem[{{Price-Whelan} {et~al.}(2018){Price-Whelan}, {Sip{\H{o}}cz},
  {G{\"u}nther}, {Lim}, {Crawford}, {Conseil}, {Shupe}, {Craig}, {Dencheva},
  {Ginsburg}, {VanderPlas}, {Bradley}, {P{\'e}rez-Su{\'a}rez}, {de Val-Borro},
  {Paper Contributors}, {Aldcroft}, {Cruz}, {Robitaille}, {Tollerud},
  {Coordination Committee}, {Ardelean}, {Babej}, {Bach}, {Bachetti}, {Bakanov},
  {Bamford}, {Barentsen}, {Barmby}, {Baumbach}, {Berry}, {Biscani}, {Boquien},
  {Bostroem}, {Bouma}, {Brammer}, {Bray}, {Breytenbach}, {Buddelmeijer},
  {Burke}, {Calderone}, {Cano Rodr{\'\i}guez}, {Cara}, {Cardoso}, {Cheedella},
  {Copin}, {Corrales}, {Crichton}, {D{\textquoteright}Avella}, {Deil},
  {Depagne}, {Dietrich}, {Donath}, {Droettboom}, {Earl}, {Erben}, {Fabbro},
  {Ferreira}, {Finethy}, {Fox}, {Garrison}, {Gibbons}, {Goldstein}, {Gommers},
  {Greco}, {Greenfield}, {Groener}, {Grollier}, {Hagen}, {Hirst}, {Homeier},
  {Horton}, {Hosseinzadeh}, {Hu}, {Hunkeler}, {Ivezi{\'c}}, {Jain}, {Jenness},
  {Kanarek}, {Kendrew}, {Kern}, {Kerzendorf}, {Khvalko}, {King}, {Kirkby},
  {Kulkarni}, {Kumar}, {Lee}, {Lenz}, {Littlefair}, {Ma}, {Macleod},
  {Mastropietro}, {McCully}, {Montagnac}, {Morris}, {Mueller}, {Mumford},
  {Muna}, {Murphy}, {Nelson}, {Nguyen}, {Ninan}, {N{\"o}the}, {Ogaz}, {Oh},
  {Parejko}, {Parley}, {Pascual}, {Patil}, {Patil}, {Plunkett}, {Prochaska},
  {Rastogi}, {Reddy Janga}, {Sabater}, {Sakurikar}, {Seifert}, {Sherbert},
  {Sherwood-Taylor}, {Shih}, {Sick}, {Silbiger}, {Singanamalla}, {Singer},
  {Sladen}, {Sooley}, {Sornarajah}, {Streicher}, {Teuben}, {Thomas},
  {Tremblay}, {Turner}, {Terr{\'o}n}, {van Kerkwijk}, {de la Vega}, {Watkins},
  {Weaver}, {Whitmore}, {Woillez}, {Zabalza}, \& {Contributors}}]{astropy2018}
{Price-Whelan}, A.~M., {Sip{\H{o}}cz}, B.~M., {G{\"u}nther}, H.~M., {et~al.}
  2018, \aj, 156, 123, \dodoi{10.3847/1538-3881/aabc4f}

\bibitem[{{Pu} \& {Lai}(2019)}]{pu2019}
{Pu}, B., \& {Lai}, D. 2019, \mnras, 488, 3568, \dodoi{10.1093/mnras/stz1817}

\bibitem[{Salvatier {et~al.}(2016)Salvatier, Wiecki, \& Fonnesbeck}]{pymc3}
Salvatier, J., Wiecki, T.~V., \& Fonnesbeck, C. 2016, PeerJ Computer Science,
  2, e55

\bibitem[{Schwarz(1978)}]{schwarz1978}
Schwarz, G. 1978, The Annals of Statistics, 6, 461,
  \dodoi{10.1214/aos/1176344136}

\bibitem[{Shporer \& Brown(2011)}]{shporer2011}
Shporer, A., \& Brown, T. 2011, The Astrophysical Journal, 733, 30,
  \dodoi{10.1088/0004-637X/733/1/30}

\bibitem[{{Steffen} \& {Coughlin}(2016)}]{steffen2016}
{Steffen}, J.~H., \& {Coughlin}, J.~L. 2016, Proceedings of the National
  Academy of Science, 113, 12023, \dodoi{10.1073/pnas.1606658113}

\bibitem[{{Theano Development Team}(2016)}]{theano}
{Theano Development Team}. 2016, arXiv e-prints, abs/1605.02688

\bibitem[{{Triaud}(2018)}]{triaud2018}
{Triaud}, A. H.~M.~J. 2018, in Handbook of Exoplanets, ed. H.~J. {Deeg} \&
  J.~A. {Belmonte}, 2

\bibitem[{{Valenti} \& {Fischer}(2005)}]{valenti2005}
{Valenti}, J.~A., \& {Fischer}, D.~A. 2005, \apjs, 159, 141,
  \dodoi{10.1086/430500}

\bibitem[{{van der Walt} {et~al.}(2011){van der Walt}, {Colbert}, \&
  {Varoquaux}}]{numpy2}
{van der Walt}, S., {Colbert}, S.~C., \& {Varoquaux}, G. 2011, Computing in
  Science Engineering, 13, 22, \dodoi{10.1109/MCSE.2011.37}

\bibitem[{{Virtanen} {et~al.}(2020){Virtanen}, {Gommers}, {Oliphant},
  {Haberland}, {Reddy}, {Cournapeau}, {Burovski}, {Peterson}, {Weckesser},
  {Bright}, {van der Walt}, {Brett}, {Wilson}, {Jarrod Millman}, {Mayorov},
  {Nelson}, {Jones}, {Kern}, {Larson}, {Carey}, {Polat}, {Feng}, {Moore}, {Vand
  erPlas}, {Laxalde}, {Perktold}, {Cimrman}, {Henriksen}, {Quintero}, {Harris},
  {Archibald}, {Ribeiro}, {Pedregosa}, {van Mulbregt}, \&
  {Contributors}}]{scipy}
{Virtanen}, P., {Gommers}, R., {Oliphant}, T.~E., {et~al.} 2020, Nature
  Methods, 17, 261, \dodoi{https://doi.org/10.1038/s41592-019-0686-2}

\bibitem[{{von Braun} {et~al.}(2011){von Braun}, {Boyajian}, {ten Brummelaar},
  {Kane}, {van Belle}, {Ciardi}, {Raymond}, {L{\'o}pez-Morales}, {McAlister},
  {Schaefer}, {Ridgway}, {Sturmann}, {Sturmann}, {White}, {Turner},
  {Farrington}, \& {Goldfinger}}]{vonbraun2011}
{von Braun}, K., {Boyajian}, T.~S., {ten Brummelaar}, T.~A., {et~al.} 2011,
  \apj, 740, 49, \dodoi{10.1088/0004-637X/740/1/49}

\bibitem[{{Winn} {et~al.}(2011){Winn}, {Matthews}, {Dawson}, {Fabrycky},
  {Holman}, {Kallinger}, {Kuschnig}, {Sasselov}, {Dragomir}, {Guenther},
  {Moffat}, {Rowe}, {Rucinski}, \& {Weiss}}]{winn2011}
{Winn}, J.~N., {Matthews}, J.~M., {Dawson}, R.~I., {et~al.} 2011, \apjl, 737,
  L18, \dodoi{10.1088/2041-8205/737/1/L18}

\bibitem[{{Zhao} {et~al.}(2022){Zhao}, {Fischer}, {Ford}, {Wise}, {Cretignier},
  {Aigrain}, {Barragan}, {Bedell}, {Buchhave}, {Camacho}, {Cegla},
  {Cisewski-Kehe}, {Collier Cameron}, {de Beurs}, {Dodson-Robinson},
  {Dumusque}, {Faria}, {Gilbertson}, {Haley}, {Harrell}, {Hogg}, {Holzer},
  {John}, {Klein}, {Lafarga}, {Lienhard}, {Maguire-Rajpaul}, {Mortier},
  {Nicholson}, {Palumbo}, {Ramirez Delgado}, {Shallue}, {Vanderburg}, {Viana},
  {Zhao}, {Zicher}, {Cabot}, {Henry}, {Roettenbacher}, {Brewer}, {Llama},
  {Petersburg}, \& {Szymkowiak}}]{zhao2022}
{Zhao}, L.~L., {Fischer}, D.~A., {Ford}, E.~B., {et~al.} 2022, \aj, 163, 171,
  \dodoi{10.3847/1538-3881/ac5176}

\end{thebibliography}

\end{document}